\documentclass[aps,twocolumn,prb]{revtex4}

\def\um{\mu\mbox{m}}

\def\Wcm2{\mbox{W cm}^{-2}}
\def\cm3{\mbox{cm}^{-3}}

\usepackage{graphicx,color}
\usepackage{amssymb}

\begin{document}

\title{Features of ion acceleration by circularly polarized laser pulses}

\author{T. V. Liseykina}
\affiliation{Dipartimento  di Fisica ``E. Fermi'', Universit\`a di Pisa, Pisa, Italy}\thanks{On leave from Institute for Computational Technologies, SD-RAS, Novosibirsk, Russia}

\author{A. Macchi}
\affiliation{polyLAB, CNR-INFM, Universit\`a di Pisa, Pisa, Italy}\email{macchi@df.unipi.it}

\date{\today}

\begin{abstract}
The characteristics of a MeV ion source driven by superintense, ultrashort
laser pulses with circular polarization are studied by means of 
particle-in-cell simulations. Predicted features include high efficiency,
large ion density, low divergence and the possibility of 
femtosecond duration. A comparison with the case of linearly polarized 
pulses is made.
\end{abstract}

\maketitle

The short-duration, multi-MeV ion beams produced in the interaction 
of high-intensity laser pulses with solid targets have proven 
to be effective for applications such as 
proton radiography,\cite{cobbleJAP02,mackinnonAPL03}
diagnostic of highly transient 
electromagnetic fields,\cite{mackinnonAPL03,borghesiAPL03,liPRL06}
isochoric heating of matter,\cite{patelPRL03} isotope production 
\cite{nemotoAPL01} and nuclear activation.\cite{mckennaAPL03} 
Foreseen future applications to medicine,\cite{malkaMP04} 
nuclear fusion \cite{atzeniNF02} or particle
physics \cite{terranovaNIMA06} will require improvements in factors such 
as the conversion efficiency, peak ion energy, beam monochromaticity and 
collimation. Recent experiments performed with these
aims \cite{hegelichN06,schwoererN06,toncianS06} are based on the 
target normal sheath acceleration (TNSA) mechanism,\cite{wilksPP01} 
where ions on a surface layer at the rear side of the target are 
accelerated by the space-charge field of escaping ``fast'' electrons.
Numerical simulations have also explored different regimes,
such as ``shock acceleration'',\cite{silvaPRL04}
``laser-piston'',\cite{esirkepovPRL04}
``skin-layer ponderomotive acceleration'',\cite{badziakAPL06}
or acceleration by circularly polarized laser pulses,\cite{macchiPRL05}
where as a common feature ion acceleration occurs at the target front side 
and is in principle dominated by the effect of the radiation pressure of 
the laser pulse. These regimes might be the leading ones at ultra-high 
intensities \cite{esirkepovPRL04} or be most suitable for specific
applications.\cite{badziakAPL06}

Here we report a numerical study on ion acceleration
with circularly polarized pulses (CP) 
with the aim to show the peculiar features of the ion source 
(high efficiency, large ion density, short duration, good collimation) 
which may be advantageous for specific applications. 
A comparison with the case of linearly polarized pulses (LP) is made
to evidentiate the differences with CP and to provide a deeper 
understanding of ``ponderomotive'' (i.e. radiation pressure--dominated)
mechanisms. This is possible because using CP at normal 
incidence fast electron generation is almost suppressed,\cite{macchiPRL05} 
thus related effects (such as TNSA) 
can be separated by purely ponderomotive ones.

We compare two one-dimensional (1D) particle-in-cell (PIC) 
simulations performed for LP and CP, respectively, 
and having same pulse wavelength ($\lambda_L=1~\mu\mbox{m}$),
duration ($\tau_L=26T_L=86~\mbox{fs}$ where $T_L=\lambda_L/c$), 
and intensity $I=3.5 \times 10^{20}~\Wcm2$.
To ensure that $I$ is the same, 
the peak field amplitude in the CP case, $a_L=11.3$, 
is lower by a factor $\sqrt{2}$ than in the LP case $a_L=16$.
Here, $a_L$ is the dimensionless pulse amplitude 
given by $a_L=0.85\sqrt{(I\lambda^2)_{18}/\alpha}$ where 
$(I\lambda^2)_{18}$ is the irradiance in units of $10^{18}~\mbox{W cm}^{-2}$
and $\alpha=1$ ($2$) for LP (CP).
The plasma parameters are the same for both simulations 
(proton plasma slab of $20~\mu\mbox{m}$ thickness and electron 
density $n_e=10n_c$ where 
$n_c=1.1 \times 10^{21}~\mbox{cm}^{-3}$ is the cut-off density).
The parameters of the LP case are close to those of simulations reported by
Silva et al.\cite{silvaPRL04} to address ``shock'' acceleration.
In both our simulations, the temporal and spatial resolution were
given by $\Delta x=c\Delta t=\lambda_L/400$ and 32 particles per
cell were used. The high spatio-temporal resolution was necessary to
ensure convergence of the results, since very sharp gradients are generated 
during the interaction, e.g. at the ion density spiking discussed 
below and elsewhere.\cite{macchiPRL05,sentokuPP03}

\begin{figure}
\includegraphics[width=0.48\textwidth]{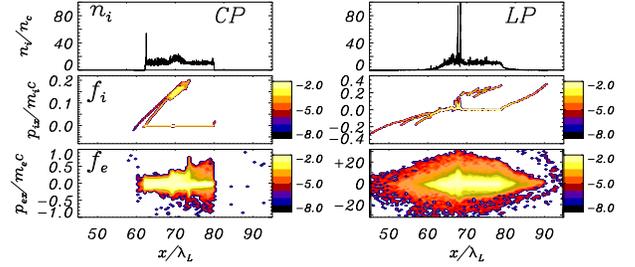}
\caption{Snaphsots at $t=140T_L=467~\mbox{fs}$
of the ion density $n_i$ (top) and the ($x,p_x$) phase
space projections of ions ($f_i$, middle) and electrons ($f_e$, bottom) from
1D PIC simulations in the CP (left) and LP (right) cases.
The laser pulse propagates from left to right.
Length scales are normalized to $\lambda_L=1~\mu\mbox{m}$,
the density to $n_c$, and momenta to $mc$. 
Notice the different scale on the momentum axes between the CP and LP cases.
\label{fig:1D}}
\end{figure}

Fig.\ref{fig:1D} compares the ion density profiles and the phase space
of ions and electrons for CP and LP. For LP, strongly relativistic 
electrons with momenta $p_{ex}$ up to $\simeq 30m_e c$ are generated.
The ion phase space shows at least three ``groups'':
ions accelerated by TNSA both at the front and the rear side, 
with momenta $p_{ix}$ up to $\simeq 0.3m_i c$,
and ions accelerated at the front surface propagating into the plasma
with similar momentum values. 
For CP, electrons are relatively ``cold'', since typical momenta are 
more than one order of magnitude lower than for LP. No significant
TNSA ions are observed, and most of the 
accelerated ions are located in a bunch with longitudinal momentum 
$p_x \simeq 0.15m_i c$.

\begin{figure}
\includegraphics[width=0.48\textwidth]{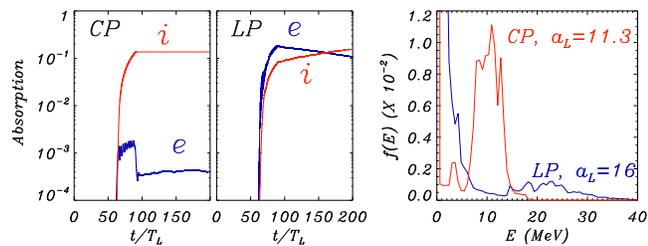}
\caption{Comparison of 
absorption efficiency into ions and electrons vs. time (left) 
and of ion energy spectra (right) 
from 1D PIC simulations of ion acceleration with 
LP or CP pulses for the same plasma parameters and laser energy and duration.
\label{fig:spectra}}
\end{figure}
  
Fig.\ref{fig:spectra} compares the absorption efficiency and the ion spectrum 
obtained in the CP and LP cases, respectively. 
The absorption into bunch ions is $13.7\%$ for CP
and is constant after the laser pulse, confirming that all ions are
``directly'' accelerated ponderomotively; absorption into electrons is 
negligible. For LP, absorption into electrons is dominant during the
interaction with the laser pulse; later, energy transfer towards ions
occurs and the conversion efficiency into ion energy has a value
similar to CP, but including all the three ion groups observed in 
Fig.\ref{fig:1D}.
The ion spectrum for CP is relatively narrow and peaked around $10~\mbox{MeV}$,
while the LP spectrum is more thermal-like, with a broad maximum 
around $20~\mbox{MeV}$. 

While ponderomotive acceleration (PA) is the only effective 
ion acceleration channel in the CP case, for LP its contribution
overlaps with TNSA, and the same analytical model proposed for the
CP case\cite{macchiPRL05} may be used if the 
longitudinal force on ions is considered to be a temporal
average. the forward accelerated ions at the front surface observed
for LP may be attributed to PA 
(rather than to ``shock'' acceleration\cite{silvaPRL04}),
and the same model 
However, the strong absorption into electrons for LP reduces the total
radiation pressure and so the PA efficiency. An estimate of the relative 
contributions of TNSA and PA is provided by an
analysis of particle energy vs. position, showing that
at the time corresponding to Fig.\ref{fig:1D}
the energy belonging to ions located within the
original plasma slab position ($60~\mu\mbox{m}<x<80~\mu\mbox{m}$)
is 76\% of the total energy for LP (with 10\% and 14\% being the contributions 
of ions emitted from the front and rear sides, respectively) and almost 
100\% for CP. 

The reason why the ``ponderomotive'' peak in the ion energy spectrum is
much more prominent for CP than for LP is attributed to the fact that the 
ponderomotive force tends to focus the ion spatially at the end of the 
skin layer, creating a very sharp density peak.\cite{macchiPRL05,sentokuPP03} 
For LP, the strong electron pressure
counteracts the piling up of the ions, leading to ``explosion'' 
of the proton bunch and to a broader energy spectrum; the maximum ion
energy is higher for LP than for CP, as a few of the ponderomotively
accelerated ions gain additional energy from the fast electrons.
The comparison with the CP simulation shows that in this latter case it is
the relatively low electron temperature which allows
for a narrow ion energy spectrum.

\begin{figure}
\includegraphics[width=0.48\textwidth]{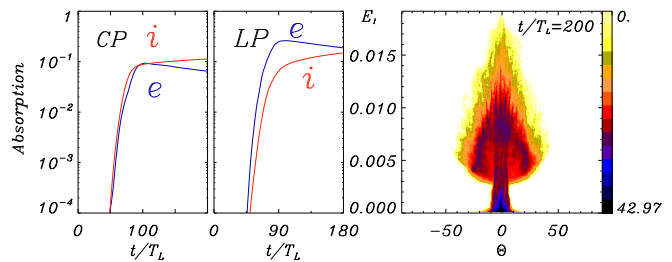}
\caption{2D simulation results. Left: comparison of 
absorption efficiency into ions and electrons vs. time.
Right: energy vs. angle distribution of ions.
The pulse radius is $r_L=2~\mu\mbox{m}$. 
Other parameters are the same of the 
1D runs of Figs.\ref{fig:1D}-\ref{fig:spectra}.
\label{fig:abs2D}}
\end{figure}

The comparison of 1D, plane-wave simulations best enlights the different
regimes of ion acceleration between LP and CP and the particular features
of PA vs. TNSA. For a realistic laser pulse with a finite spot size, 
the differences between LP and CP are somewhat weakened by the effects of
pulse focusing which introduces electric field components normal to 
the target surface at the edges of the spot, leading to electron heating.
Fig.\ref{fig:abs2D} shows the effect on fractional absorption in 2D
simulations with the same parameters of the 1D case, and a tightly focused
pulse with a spot radius $r_L=2\lambda_L$. The differences between LP
and CP and between ion and electron absorption in the latter case are less
dramatic as expected, but still substantial.

The angular spread of ions depends upon their energy $E$. 
The $f(E,\Theta)$ distribution of ions in  [Fig.\ref{fig:abs2D} b)], 
where $\Theta$ is the emission angle with respect to the target normal
shows that the most energetic the ions
the most collimated they are: for instance, ions having energy
exceeding $0.01m_i c^2$ are found within a cone with an aperture angle 
of about 10 degrees.
Wider spot sizes or smoother intensity profiles may yield a lower 
divergence.

The narrow ion energy spectrum of the CP case is a necessary condition to 
obtain a dense ion bunch with very short duration, in addition to the 
requirement of a laser pulse with duration of the order of the bunch
acceleration time.\cite{macchiPRL05} These particular features of ions
accelerated using CP pulses may be useful for ultrafast, 
localized energy deposition in matter, 
and are essential for a proposed concept of sources of fusion neutrons 
with duration of a few femtoseconds.\cite{macchiAPB06}

\begin{figure}
\includegraphics[width=0.48\textwidth]{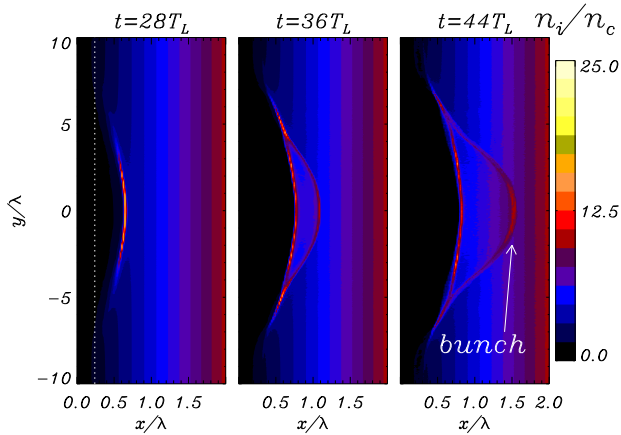}
\caption{Snapshots of ion density from a 2D, CP simulation 
for a linear density profile rising from $0$ to $10n_c$ over $2\lambda$.
Laser pulse parameters are $a_L=2$, $\tau_L=10T_L$ and $r_L=4\lambda$.
The white dotted line indicates the initial position of the $n_e=n_c$ surface.
The arrow indicates the location of the short ion ``bunch'' at $t=44T_L$. 
\label{fig4}}
\end{figure}

The production of a single, ultrashort ion bunch can be observed in the
simulation of Fig.\ref{fig4} (for which $I=5.5\times 10^{18}~\Wcm2$,
$\tau_L=33~\mbox{fs}$ and $r_L=4~\um$) 
where, in addition, a linear density profile
was used instead of a step-like one to address the effect of early plasma
production by a prepulse in experiments. We observe that ion acceleration
initially occurs near the cut-off layer where $n_e=n_c$, and produces at
$t \simeq 30T_L$ a narrow ion bunch with a density larger than $10n_c$.
The divergence of the bunch ions is about 4 degrees.

So far, laser-plasma interaction experiments with CP have been rarely 
reported. Kado et al.\cite{kadoLPB06} observed
a collimated proton beam in the interaction of 
CP pulses with plastic-coated Tantalum targets 
at an intensity of $4\times 10^{18}~\mbox{W cm}^{-2}$.
In addition, no electrons with energy beyond $20~\mbox{keV}$ were
observed in these conditions.\cite{daidoPC} 
These observations are in qualitative agreement with the above scenario 
of CP laser ion acceleration. 
Fukumi et al.\cite{fukumiPP05} also reported
ion energy measurements for a CP
laser pulse but, since a substantial incidence angle
($45$ deg) was used, the interaction conditions were rather similar to an
overlap of $s$- and $p$-polarized laser pulses, causing a dominant effect of 
electron heating and TNSA of ions. 
Since no unaffordable problem seems to prevent the use of 
high-intensity CP pulses at normal incidence for laser-plasma interactions,
the regime of ion acceleration with CP pulses may be investigated in 
present-day experiments allowing progress in the use of laser-accelerated
ions for specific applications.

This work has been supported 
by the Italian Ministery for University and Research 
(MIUR) via the PRIN project ``Ultraintense laser-plasma interaction'', 
by CNR-INFM and CINECA (Italy) through the super-computing initiative,
and by JSCC (Moscow, Russia). 

\newpage
\hyphenation{Post-Script Sprin-ger}

\end{document}